\pdfoutput=1 
\documentclass[12pt,a4paper]{article}%
\usepackage{amsmath,amssymb,amsfonts}
\usepackage{graphicx,epsfig,url}
\usepackage{color}
\usepackage{amsmath}
\usepackage{amsfonts}
\usepackage{amssymb}
\usepackage{float}
\usepackage{caption2}
\usepackage{graphicx}
\usepackage[]{hyperref}
\usepackage{dsfont}
\numberwithin{equation}{section}
 \setcaptionmargin{1cm}

\def\beq{\begin{equation}}
\def\eeq{\end{equation}}
\def\bea{\begin{eqnarray}}
\def\eea{\end{eqnarray}}

\newcommand{\newc}{\newcommand}
\newc{\gsim}{\lower.7ex\hbox{$\;\stackrel{\textstyle>}{\sim}\;$}}
\newc{\lsim}{\lower.7ex\hbox{$\;\stackrel{\textstyle<}{\sim}\;$}}
\newc{\ie}{{\it i.e.}}
\newc{\etal}{{\it et al.}}
\newc{\mev}{\hbox{\rm\,MeV}}
\newc{\gev}{\hbox{\rm\,GeV}}
\newc{\tev}{\hbox{\rm\,TeV}}
\begin{document}


\begin{titlepage}
\begin{flushright}
\end{flushright}
\vskip 1.0cm
\begin{center}
{\Large \bf Universal Constraints on Conformal Operator Dimensions
} \vskip 1.0cm {\large
Vyacheslav S.~Rychkov$^{\,a}$\ \  and\ \  Alessandro Vichi$\,^b$
} \\[1cm]
{\it $^a$ Scuola Normale Superiore and INFN, Pisa, Italy} \\[5mm]
{\it $^b$ Institut de Th\'eorie des Ph\'enom\`enes Physiques, EPFL, Lausanne, Switzerland}\\[5mm]
\vskip 1.0cm \abstract{ We continue the study of model-independent
constraints on the unitary Conformal Field Theories in 4-Dimensions,
initiated in \href{http://arxiv.org/abs/0807.0004}{arXiv:0807.0004}.
Our main result is an improved upper bound on the dimension $\Delta$
of the leading scalar operator appearing in the OPE of two identical
scalars of dimension $d$: $$\phi_d \times \phi_d={\mathds
1}+O_{\Delta}+\ldots$$ In the interval $1<d<1.7$ this universal
bound takes the form
$$
\Delta \leq2+0.7(d-1)^{1/2}+2.1(d-1)+0.43(d-1)^{3/2}.$$ The proof is
based on prime principles of CFT: unitarity, crossing symmetry, OPE,
and conformal block decomposition. We also discuss possible
applications to particle phenomenology and, via a 2-D analogue, to
string theory. }
\end{center}
\end{titlepage}

\section{Introduction and formulation of the problem}

\label{sec:intro}

Our knowledge about non-supersymmetric Conformal Field Theories (CFTs) in four
dimensions (4D) is still quite incomplete. Suffices it to say that not a
single nontrivial example is known which would be solvable to the same extent
as, say, the 2D Ising model. However, we do not doubt that CFTs must be
ubiquitous. For example, non-supersymmetric gauge theories with $N_{c}$ colors
and $N_{f}$ flavors are widely believed to have \textquotedblleft conformal
windows\textquotedblright\ in which the theory has a conformal fixed point in
the IR, with evidence from large $N_{c}$ analysis \cite{belavin},
supersymmetric analogues \cite{seiberg}, and lattice simulations
\cite{lattice}. Since these fixed points are typically strongly coupled, we do
not have much control over them. In this situation particularly important are
general, model-independent properties.

One example of such a property is the famous unitarity bound \cite{mack} on
the dimension $\Delta$ of a spin $l$ conformal primary operator $O_{\Delta,l}$
:\footnote{Here we quote only the case of symmetric traceless tensor
operators.}%
\begin{align}
&  \Delta\geq1\quad(l=0)\,,\label{eq-unit}\\
&  \Delta\geq l+2\quad(l\geq1)\,.\nonumber
\end{align}
These bounds are derived by imposing that the two point function $\left\langle
OO\right\rangle $ have a positive spectral density.

\begin{figure}
[ptbh]
\begin{center}
\includegraphics [height=0.9772in] {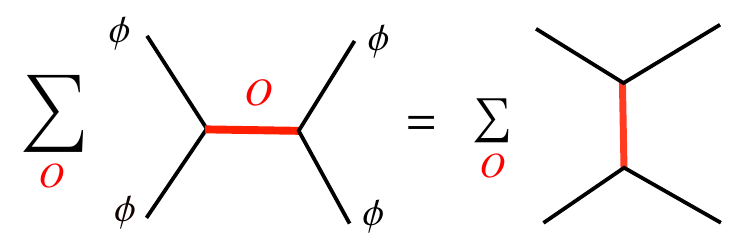}%
\caption{\emph{The conformal bootstrap equation. The thick red line
denotes a conformal block, summing up exchanges of a primary
operator }$O$\emph{ and all
its descendants.}}%
\label{fig:bootstrap}%
\end{center}
\end{figure}

As is well known, 3-point functions in CFT are fixed by conformal symmetry up
to a few arbitrary constants (Operator Product Expansion (OPE)\ coefficients).
The next nontrivial constraint thus appears at the 4-point function level, and
is known as the \textit{conformal bootstrap} equation. It says that OPE
applied in direct and crossed channel should give the same result (see
Fig.~\ref{fig:bootstrap}).

The bootstrap equation goes back to the early days of CFT
\cite{polyakov}. However, until recently, not much useful general
information has been extracted from it\footnote{Except in 2D, in
theories with finitely many primary fields and in the Liouville
theory \cite{ZZ}. We will comment on the 2D case in Sections
\ref{sec:2Dpreview} and \ref{sec:outlook} below.}. All spins and
dimensions can apriori enter the bootstrap on equal footing, and
this seems to lead to unsurmountable difficulties.%

Recently, however, tangible progress in the analysis of bootstrap equations
was achieved in \cite{R}. Namely, it was found that, in unitary theories, the
functions entering the bootstrap equations (conformal blocks) satisfy certain
\textit{positivity properties} which lead to general necessary conditions for
the existence of solutions.

The concrete problem considered in \cite{R}, and which we will continue to
discuss here, was as follows. In an arbitrary unitary CFT a Hermitean scalar
primary $\phi_{d}$ of dimension $d$ was singled out. The conformal bootstrap
equation for its 4-point function $\left\langle \phi_{d}\phi_{d}\phi_{d}%
\phi_{d}\right\rangle $ was studied under the sole assumption that all
\emph{scalars} in the OPE $\phi_{d}\times\phi_{d}$ have dimension above a
certain number, call it $\Delta_{\min}:$%
\begin{equation}
\phi_{d}\times\phi_{d}=\mathds{1}~\text{+~\thinspace(Scalars of dimension}%
\geq\Delta_{\min})~\,\text{+~\thinspace(Higher spins)\thinspace.}
\label{eq-ope0}%
\end{equation}
It was shown that the conformal bootstrap \textit{does not allow for a
solution} unless%
\begin{equation}
\Delta_{\min}\leq f(d)\,, \label{eq-bound}%
\end{equation}
where $f(d)$ is a certain continuous function, computed numerically. We stress
that this conclusion was reached without making any assumptions about
dimensions or spins of other operators appearing in the OPE, beyond those
implied by the unitarity bounds. Nor any assumptions about the OPE
coefficients were made (apart from their reality, which is again implied by unitarity).

In other words, \textit{in \textbf{any} unitary 4D CFT, the OPE of
\textbf{any} scalar primary }$\phi_{d}$\textit{ must contain at least one
scalar field }$O_{\Delta}$\textit{ with dimension not larger than }$f(d).$

Incidentally, the function $f(d)$ was found to satisfy $f(1)=2,$ which is
quite natural since $d=1$ corresponds to the free field whose OPE contains the
operator :$\phi^{2}$: of dimension $2$.

What makes the result like (\ref{eq-bound}) possible? The basic reason is
that, in any theory, crossing symmetry relation of Fig.~\ref{fig:bootstrap}
cannot be satisfied term by term, but only by cancellations among various
terms. The guaranteed presence of the unit operator in the OPE (\ref{eq-ope0})
creates a certain \textquotedblleft crossing symmetry
deficit\textquotedblright, which has to be balanced by other fields. The idea
is to show that this cannot happen unless at least one scalar of sufficiently
low dimension is present.%
\begin{figure}
[t]
\begin{center}
\includegraphics[height=0.7922in,width=3.8847in]{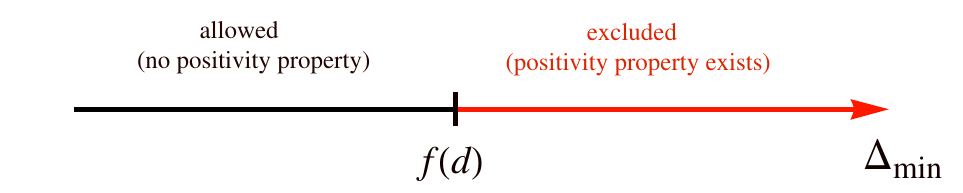}%
\caption{\emph{The bound }$f(d)$\emph{ is the smallest }$\Delta_{\min}$\emph{
for which a positivity property exists. }}%
\label{figDichotomy}%
\end{center}
\end{figure}

Technically, the method of \cite{R} consists of 3 steps (see Section
\ref{sec:positive} for a detailed review):

\begin{enumerate}
\item We Taylor-expand the conformal bootstrap equation near the
\textquotedblleft self-dual point" configuration having equal conformal
cross-ratios $u=v$. The expansion is truncated to a certain finite order $N$.

\item We systematically search for \textit{positivity properties} satisfied by
linear combinations of Taylor coefficients of the conformal blocks, for fields
appearing in the RHS of the OPE (\ref{eq-ope0}). A found positivity property
implies that the \textquotedblleft crossing symmetry deficit\textquotedblright%
\ cannot be balanced and rules out a CFT with a given $d$ and $\Delta_{\min}$.

\item For fixed $d$, the bound $f(d)$ is then computed as the point separating
those $\Delta_{\min}$ for which a positivity property exists, from those ones
for which it does not (Fig.~\ref{figDichotomy}).
\end{enumerate}

\begin{figure}
[ptbh]
\begin{center}
\includegraphics[
height=2.341in,
width=3.3797in
]%
{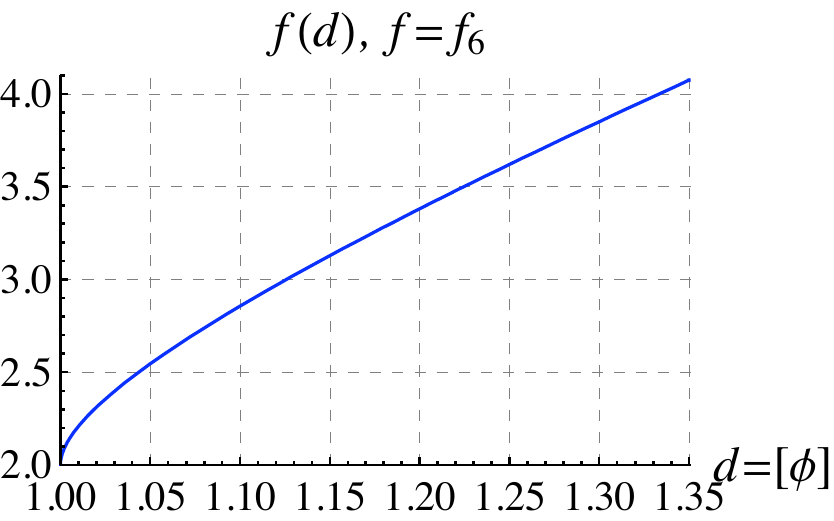}%
\caption{\emph{The bound }$f_{6}(d)\simeq2+1.79\sqrt{d-1}+2.9(d-1)$, $1\leq
d\leq1.35$,\emph{ corresponding to }$N=6$\emph{, reproduced from \cite{R}.}}%
\label{figR}%
\end{center}
\end{figure}

The nature of the method is such that increasing $N$ can make the bound only
stronger.The optimal bound should in principle be recoverable in the limit
$N\rightarrow\infty$. In practice the value of $N$ is determined by the
available computer resources and algorithmic efficiency. The best bound found
in \cite{R}, plotted in Fig.~\ref{figR}, corresponds to $N=6.$%

The purpose of this paper is to present an improvement of the bound
(\ref{eq-bound}) obtained by using the method of \cite{R} with larger values
of $N$, up to $N=18$. The new results are interesting in two ways. First, pure
numerical improvement turns out to be significant. Second, $N=18$ happens to
be large enough so that we start observing saturation of the bound. So we
believe our current results are close to the optimal ones achievable with this method.

The paper is organized as follows. In Section \ref{sec:review} we review the
conformal bootstrap equations. In Section \ref{sec:positive} we review the
connection of the bound (\ref{eq-bound}).with positivity properties satisfied
by the conformal block expansion coefficients. In Section \ref{sec:results} we
present and discuss our results. We also mention accompanying results which we
obtain for an analogous problem in 2D. In Section \ref{sec:outlook} we propose
several future applications and extensions of our method, with emphasis on
connections to phenomenology and string theory. In Section \ref{sec:summary}
we summarize and conclude. In Appendix \ref{sec:numerics} we collect some
details about our numerical algorithms. In Appendix \ref{sec:tables} we
include the tables on which plots in Section \ref{sec:results} are based.

\section{Review of conformal bootstrap}

\label{sec:review}

We will review the conformal bootstrap equation in its simplest form---as
applied to the 4-point function of identical scalars $\left\langle \phi
\phi\phi\phi\right\rangle $. We largely follow \cite{R}, where a more detailed
discussion and references can be found.

\subsection{Conformal block decomposition}

\label{sec:decomposition}

Let $\phi\equiv\phi_{d}$ be a Hermitean scalar primary\footnote{The field is
called primary if it transforms homogeneously under the $4D$ conformal group.}
operator. The operator product expansion (OPE) $\phi\times\phi$ contains, in
general, infinitely many primary fields of arbitrary high spins and
dimensions:\footnote{If there are several primaries with the same $\Delta,l$,
they have to be all included in this sum with independent coefficients.}%
\begin{equation}
\phi(x)\phi(0)\sim\frac{1}{|x|^{2d}}\left\{  \mathds{1}+\sum_{l=2n}%
c_{\Delta,l}\left[  \vphantom{\sum}|x|^{-\Delta}K_{l}(x)\cdot O_{\Delta
,l}(0)+\cdots\right]  \right\}  \,,\quad K_{l}(x)=\frac{x^{\mu_{1}}\cdots
x^{\mu_{l}}}{|x|^{l}}\,. \label{eq-OPE}%
\end{equation}
Here

\begin{itemize}
\item $l=2n$ by Bose symmetry;

\item $\Delta\geq1$ $(\Delta\geq l+2)$\ for $l=0$ ($l\geq2$) by the unitarity
bounds (\ref{eq-unit});

\item The $\cdots$\ stands for contributions of descendants of the primary
$O_{\Delta,l}$ (i.e.~its derivatives). These contributions are fixed by
conformal symmetry;

\item The OPE coefficients $c_{\Delta,l}$ are real (see Appendix A of \cite{R}).
\end{itemize}

\begin{figure}
[t]
\begin{center}
\includegraphics[
height=1.5342in,
width=2.4837in
]%
{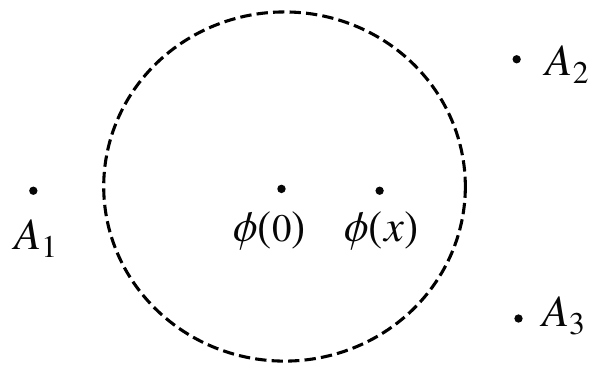}%
\caption{\emph{The operator product expansion of }$\phi(x)\phi(0)$\emph{
converges for this configuration. }}%
\label{figOPE}%
\end{center}
\end{figure}

We assume that the OPE converges in the following weak sense:~it gives a
convergent power series expansion for any $(2+n)$-point function%
\[
\left\langle \phi(x)\phi(0)A_{1}(y_{1})\ldots A_{n}(y_{n})\right\rangle
\]
provided that $\left\vert x\right\vert <\left\vert y_{i}\right\vert $, i.e.
$\phi(x)$ is closer to the origin than any other local field insertion (see
Fig.~\ref{figOPE}). This assumption can be justified by using radial
quantization (\cite{pol1}, Sect.~2.9), and checked explicitly in free field
theory. For rigorous mathematical results about OPE convergence see
\cite{mackOPE}.%

The OPE (\ref{eq-OPE}) can be used to obtain \emph{conformal block
decomposition} of the 4-point function $\left\langle \phi\phi\phi
\phi\right\rangle $:%
\begin{align}
&  \left\langle \phi(x_{1})\,\phi(x_{2})\,\phi(x_{3})\,\phi(x_{4}%
)\right\rangle =\frac{g(u,v)}{x_{12}^{2d}\,x_{34}^{2d}}\,,\label{eq-4pt}\\
&  g(u,v)=1+\sum p_{\Delta,l}\,g_{\Delta,l}(u,v)\,,\quad p_{\Delta,l}\equiv
c_{\Delta,l}^{2}\geq0, \label{eq-guv}%
\end{align}
where $u=x_{12}^{2}x_{34}^{2}/(x_{13}^{2}x_{24}^{2}),$ $v=x_{14}^{2}x_{23}%
^{2}/(x_{13}^{2}x_{24}^{2})$ are the conformal cross-ratios. This
representation is obtained by using the OPE in the 12 and 34 channels. The
\emph{conformal blocks} $g_{\Delta,l}(u,v)$ sum up the contributions of the
primary $O_{\Delta,l}$ and all its descendants. Their explicit expressions
were found by Dolan and Osborn \cite{do1}:%
\begin{align}
&  g_{\Delta,l}(u,v)=\frac{(-)^{l}}{2^{l}}\frac{z\bar{z}}{z-\bar{z}}\left[
\,k_{\Delta+l}(z)k_{\Delta-l-2}(\bar{z})-(z\leftrightarrow\bar{z})\right]
\,,\nonumber\\[0.14in]
&  \qquad\qquad k_{\beta}(x)\equiv x^{\beta/2}{}_{2}F_{1}\left(  \beta
/2,\beta/2,\beta;x\right)  \,,\label{eq-DO}\\[0.14in]
&  \qquad\quad\qquad u=z\bar{z},\quad v=(1-z)(1-\bar{z})\,.\nonumber
\end{align}

\begin{figure}
[ptb]
\begin{center}
\includegraphics[
height=1.3612in,
width=2.8323in
]%
{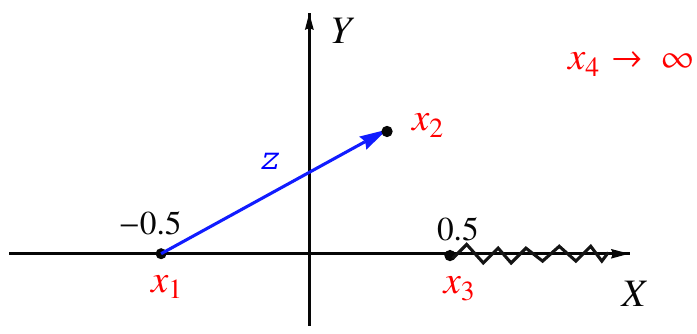}%
\caption{\emph{The auxiliary }$z$\emph{ coordinate. The conformal blocks are
regular outside the cut denoted by the zigzag line.}}%
\label{figZ}%
\end{center}
\end{figure}

Notice the judicious introduction of the auxiliary variables $z$ and $\bar
{z}.$ When the theory is formulated in the Euclidean space, these variables
are complex-conjugates of each other. To understand their meaning, it is
convenient to use the conformal group freedom to send $x_{4}\rightarrow\infty$
and to put the other three points in a plane, as in Fig.~\ref{figZ}. Then it's
easy to show that%
\begin{equation}
z=\frac{1}{2}+X+iY,\quad\bar{z}=z^{\ast}\,, \label{eq-zzbar}%
\end{equation}
where $(X,Y)$ are the coordinates of $x_{2}$ in the plane, chosen so that
$X=Y=0$ corresponds to $x_{2}$ halfway between $x_{1}$ and $x_{3}.$ This
\textquotedblleft self-dual\textquotedblright\ configuration, for which $u=v$,
will play an important role below. We can see that the $z$ variable is a
natural extension of the usual complex coordinate of the 2D CFT to the 4D case.%

According to the above discussion, the OPE is expected to converge for
$|z|<1.$ Conformal block decomposition is a partial resummation of the OPE and
thus also converges at least in this range. In fact, below we will only use
convergence around the self-dual point $z=1/2$. However, conformal blocks, as
given by (\ref{eq-DO}), are regular (real-analytic) in a larger region, namely
in the $z$-plane with the $\left(  1,+\infty\right)  $ cut along the real axis
(see Fig.~\ref{figZ}). The conformal block decomposition is thus expected to
converge in this larger region. One can check that this indeed happens in the
free scalar theory.

One can intuitively understand the reason for this extended region of
regularity. The condition for the OPE convergence, as stated above, does not
treat the points $x$ and $0$ symmetrically. On the other hand, the conformal
blocks are completely symmetric in $x_{1}\leftrightarrow x_{2}$ and so must be
the condition for their regularity. The appropriate condition is as follows:
\emph{the conformal block decomposition in the }$12$\emph{-}$34$\emph{ channel
is regular and convergent if there is a sphere separating the points }%
$x_{1,2}$\emph{ from the points }$x_{3,4}.$ For the configuration of
Fig.~\ref{figZ}, such a sphere exists as long as $x_{2}$ is away from the cut.

\subsection{Conformal bootstrap and the sum rule}

\label{sec:sumrule}

The 4-point function in (\ref{eq-4pt}) must be symmetric under the interchange
of any two $x_{i}$, and its conformal block decomposition (\ref{eq-guv}) has
to respect this symmetry. The symmetry with respect to $x_{1}\leftrightarrow
x_{2}$ or $x_{3}\leftrightarrow x_{4}$ is already built in, since only even
spins are exchanged \cite{do1}. On the contrary, the symmetry with respect to
$x_{1}\leftrightarrow x_{3}$ gives a condition%
\begin{equation}
v^{d}g(u,v)=u^{d}g(v,u)\,, \label{eq-boot}%
\end{equation}
which is not automatically satisfied for $g(u,v)$ given by (\ref{eq-guv}).
This nontrivial constraint on dimensions, spins, and OPE coefficients of all
operators appearing in the OPE $\phi\times\phi$ is known as the
\textit{conformal bootstrap equation}. Physically it means that OPE applied in
12-34 and 14-23 channels should give the same result (Fig.~\ref{fig:bootstrap}).

In the $z$-plane of Section \ref{sec:decomposition}, the LHS of (\ref{eq-boot}%
) has a cut along $(1,+\infty),$ while the RHS has a cut along $(-\infty,0).$
Thus, if (\ref{eq-boot}) is satisfied, the cuts have to cancel, and the
resulting $g(u,v)$ is real analytic everywhere except for $z=0,1.$

In \cite{R}, we found it useful to rewrite (\ref{eq-boot}) by separating the
unit operator contribution, which gives
\begin{equation}
u^{d}-v^{d}=\sum p_{\Delta,l}\left[  v^{d}g_{\Delta,l}(u,v)-u^{d}g_{\Delta
,l}(v,u)\right]  . \label{eq-1sep}%
\end{equation}
The LHS of this equation is the \textquotedblleft crossing symmetry
deficit\textquotedblright\ created by the presence of the unit operator in the
OPE. This deficit has to be balanced by contributions of the other fields in
the RHS.

In practice it is convenient to normalize (\ref{eq-1sep}) by dividing both
sides by $u^{d}-v^{d}$. The resulting \textit{sum rule} takes the form:%
\begin{align}
1  &  =\sum p_{\Delta,l}F_{d,\Delta,l}(X,Y)\,,\quad\nonumber\\
F_{d,\Delta,l}(X,Y)  &  \equiv\frac{v^{d}g_{\Delta,l}(u,v)-u^{d}g_{\Delta
,l}(v,u)\,}{u^{d}-v^{d}}. \label{eq-F}%
\end{align}
The \textquotedblleft F-functions" $F_{d,\Delta,l}$ are real and regular in
the full $z$-plane cut along $(-\infty,0)\cup(1,+\infty).$ In particular, the
$0/0$ behavior at the self-dual point $z=1/2$ is regular.

All F-functions vanish near the points $z=0$ and $z=1.$ Thus the sum rule can
never be satisfied near these points if only finitely many terms are present
in the RHS. \emph{The OPEs containing finitely many primaries are ruled out.}

\section{Positivity argument}

\label{sec:positive}

The main idea of \cite{R} was very simple, and can be described as follows.
Suppose that for a given spectrum of operator dimensions and spins $\left\{
\Delta,l\right\}  $ the sum rule (\ref{eq-F}), viewed as an equation for the
coefficients $p_{\Delta,l}\geq0$, has no solution. Then of course such a
spectrum would be ruled out.

Any concrete realization of this idea needs a practical criterium to show that
there is no solution. For a prototypical example of such a criterium, imagine
that a certain derivative, e.g. $\partial_{X}$ (see (\ref{eq-zzbar})), when
applied to every $F_{d,\Delta,l}$ and evaluated at a certain point, is
strictly positive (\textquotedblleft positivity property\textquotedblright).
Since the same derivative applied to the LHS of (\ref{eq-F}) gives identically
zero, a solution where all coefficients $p_{\Delta,l}$ are non-negative would
clearly be impossible. We refer to this simple reasoning as the
\textquotedblleft positivity argument\textquotedblright.

One can imagine more general criteria using different differential operators,
and applying them at different points. In \cite{R}, we found it convenient to
apply differential operators precisely at the self-dual point\ $z=1/2$,
$X=Y=0.$ One can show that the F-functions are even with respect to this point
both in the $X$ and $Y$ directions:%
\[
F(X,Y)=F(X,-Y)=F(-X,Y).
\]
Thus, all odd-order derivatives vanish, and a sufficiently general
differential operator (\textquotedblleft linear functional\textquotedblright)
takes the form:%
\begin{equation}
\Lambda\lbrack F]=\sum_{\substack{m,n\text{ even}\\2\leq m+n\leq N}%
}\lambda_{m,n}\,\partial_{X}^{m}\partial_{Y}^{n}F|_{X=Y=0}\,,
\label{eq-Lambda}%
\end{equation}
where $N$ is some fixed finite number, and $\lambda_{m,n}$ are fixed real
coefficients.\footnote{In \cite{R}, we analytically continued to the Minkowski
space by Wick-rotating $Y\rightarrow iT.$ In this picture $z$ and $\bar{z}$
are both real and independent, and conformal blocks are real regular functions
in the region $0<z,\bar{z}<1.$ For our purposes Minkowski and Euclidean
pictures are exactly equivalent. In particular, derivatives of F-functions in
$Y$ and $T$ are trivially proportional to each other.} Notice the exclusion of
the constant term $m=n=0$, in order to have $\Lambda\lbrack1]=0$.

\textit{Assume that for certain fixed }$d$\textit{ and }$\Delta_{\text{min}}%
,$\textit{ we manage to find a linear functional of this form such that
(\textquotedblleft positivity property\textquotedblright)}%
\begin{align}
\Lambda\lbrack F_{d,\Delta,l}]\geq0\text{ }  &  \text{for all }\Delta
\geq\Delta_{\min}\,(l=0)\label{eq-ineq}\\
\text{and }  &  \text{for all }\Delta\geq l+2\,(l=2,4,6\ldots)\,.\nonumber
\end{align}
\textit{Moreover, assume that all but a finite number of these inequalities
are actually strict: }$\Lambda\lbrack F]>0.$ \textit{Then the sum rule cannot
be satisfied, and such a spectrum, corresponding to a putative OPE
(\ref{eq-ope0}), is ruled out.}

The proof uses the above \textquotedblleft positivity
argument\textquotedblright. Since $\Lambda\lbrack1]=0,$ the positivity
property implies that only those primaries for which $\Lambda\lbrack F]=0$
would be allowed to appear in the RHS of the sum rule with nonzero
coefficients. By assumption, there are at most a finite number of such
primaries. However, as noted in Section \ref{sec:sumrule}, finitely many terms
can never satisfy the sum rule globally, because of the behavior near $z=0,1.$ Q.E.D.

While the above formal reasoning is quite sufficient to understand our
results, in \cite{R} the sum rule was also given an alternative interpretation
in terms of convex geometry. In this more visual picture, linear combinations
of F-functions with arbitrary positive coefficients form a \textit{convex
cone} in the space of two-variable functions. One can consider the full
function space or its finite-dimensional subspace corresponding to
Taylor-expanding up to order $N.$ Positivity property (\ref{eq-ineq}) means
that there is a hyperplane separating the function 1 from the convex cone.
Thus it implies that the sum rule cannot be satisfied. The converse is
\textquotedblleft almost true\textquotedblright, modulo questions of convergence.

Clearly, the language of linear functionals provides an equivalent, dual
formulation of the problem. This formulation is also especially convenient
from the point of view of checking our results independently. It's not so
important \emph{how} we find the functionals. As long as we publish the
functional coefficients $\lambda_{m,n}$, anyone can verify that the
inequalities (\ref{eq-ineq}) are satisfied.

\section{Results, discussion, and 2D analogue}

\label{sec:results}

As discussed in Section \ref{sec:intro}, we are interested in computing an
upper bound $($\ref{eq-bound}$)$ for the dimension $\Delta_{\min}$ of the
leading scalar in the OPE $\phi_{d}\times\phi_{d}$, universal for all unitary
4D CFTs. In \cite{R}, we have computed such a bound in the interval $1\leq
d\leq1.3,$ using the sum rule of Section \ref{sec:sumrule} truncated to the
$N=6$ derivative order. That bound is reproduced in Fig.~\ref{figR}.

\begin{figure}
[ptbh]
\begin{center}
\includegraphics[
height=2.9776in,
width=4.0811in
]%
{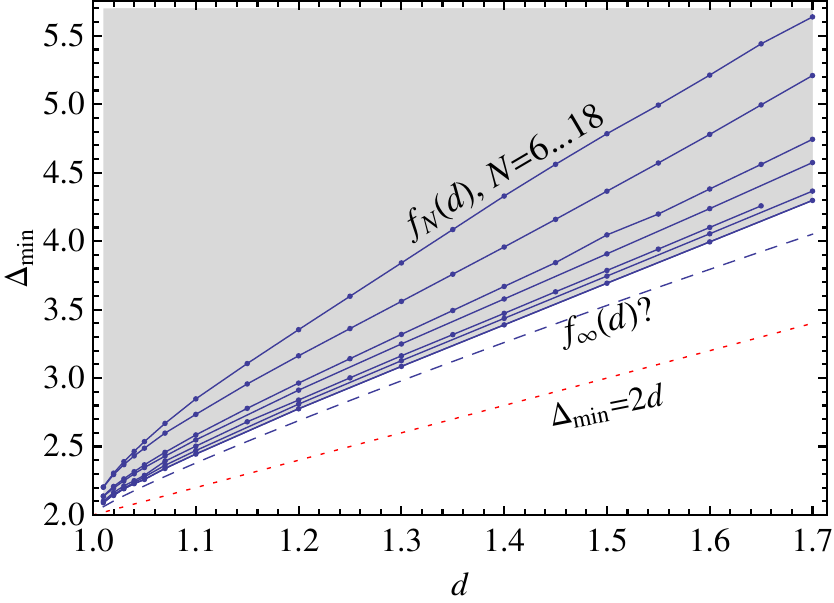}%
\caption{\textit{Our main results. The solid curves are the bounds }$f_{N}%
(d)$\textit{, }$N=6\ldots18$\textit{. The bounds get stronger as }$N$\textit{
increases, thus }$N=6$\textit{ is the weakest bound (highest curve), and
}$N=18$\textit{ is the current best bound (lowest curve). The shaded region is
thus excluded. The dashed curve }$f_{\infty}(d)$\textit{ is an approximation
to the best possible bound, obtained by extrapolating }$N\rightarrow\infty
$\textit{. The dotted line }$\Delta_{\min}=2d$\textit{ is realized in a family
of \textquotedblleft generalized free scalar\textquotedblright\ CFTs, and is
compatible with our bounds. }}%
\label{figN}%
\end{center}
\end{figure}

We now present the results of our latest study, obtained for larger values of
$N$. These results\footnote{See Appendix B for the same results in tabular
form.} are plotted in Fig \ref{figN} as a collection of curves $f_{N}(d)$,
$N=6\ldots18$, where the index $N$ denotes the number of derivatives used to
obtain the bound. The bound naturally gets stronger as $N$ increases (see
below), and thus the lowest curve $f_{18}(d)$ is the strongest bound to date.
In the considered interval $1\leq d\leq1.7$ this bound is well approximated
(within $0.5\%$) by%
\begin{equation}
f_{18}(d)\simeq2+0.7\gamma^{1/2}+2.1\gamma+0.43\gamma^{3/2},\quad\gamma=d-1.
\label{eq-f18}%
\end{equation}

To obtain the bounds of Fig.~\ref{figN}, we used the positivity argument from
\cite{R}, as reviewed in Section \ref{sec:positive}. Namely, for points lying
on the curves $\Delta_{\min}=f_{N}(d)$ we are able to find a linear functional
of the form (\ref{eq-Lambda}) satisfying the positivity property
(\ref{eq-ineq}).\footnote{Thus actually the bound is strict: $\Delta_{\min
}<f_{N}(d)$, except at $d=1.$} The numerical procedure that we use to find
these \textquotedblleft positive functionals\textquotedblright\ is described
in some detail in Appendix A.%

%

\begin{figure}
[ptbh]
\begin{center}
\includegraphics[
height=3.1695in,
width=4.088in
]%
{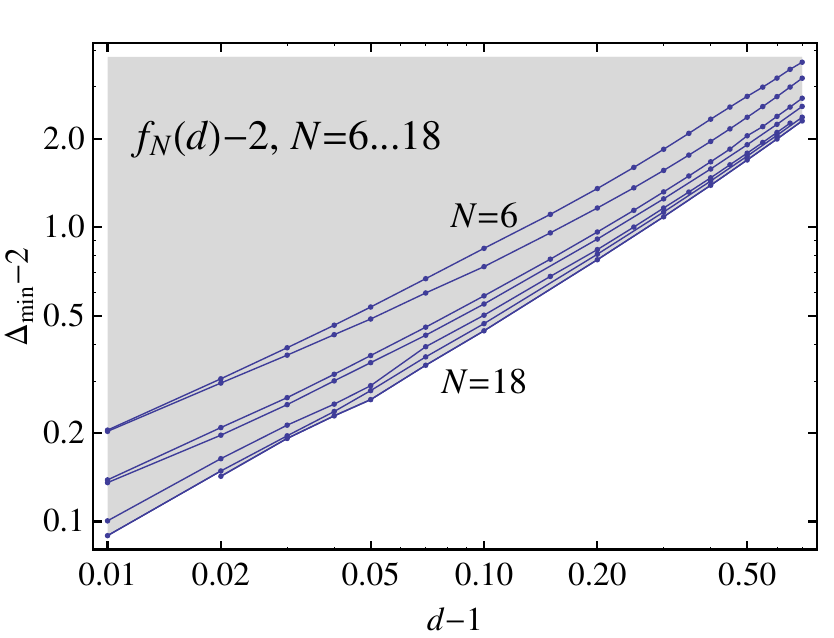}%
\caption{\textit{Same as Fig.~\ref{figN}~but with anomalous dimensions }%
$d-1,$\textit{ }$\Delta-2$\textit{ in logarithmic scale. The shaded region is
excluded.}}%
\label{figNlog}%
\end{center}
\end{figure}

Several comments are in order here.

\begin{enumerate}
\item We have actually computed the bound only for a discrete number of $d$
values, shown as points in Fig.~\ref{figN}. The tables of these computed
values are given in Appendix B. Behavior for $d\rightarrow1$ can be better
appreciated from the logarithmic-scale plot in Fig. \ref{figNlog}.

\hspace{0.5cm} We do not see any significant indication which could suggest
that the curves $f_{N}(d)$ do not interpolate smoothly in between the computed
points. Small irregularities in the slope are however visible at several
points in Figs.~\ref{figN},\ref{figNlog}. These irregularities are understood;
they originate from the necessity to \textit{discretize} the infinite system
of inequalities (\ref{eq-ineq}), see Appendix A for a discussion. In our
computations the discretization step was chosen so that these irregularities
are typically much smaller than the improvement of the bound that one gets for
$N\rightarrow N+2$.

\item For each $N$ the bound $f_{N}(d)$ is near-optimal, in the sense that no
positive functional involving derivatives up to order $N$ exists for
\[
\Delta_{\min}-2<(1-\varepsilon)[f_{N}(d)-2].
\]
We estimate $\varepsilon\simeq1\%$ from the analysis of residuals in the fit
of $f_{N}(d)$ by a smooth curve like in (\ref{eq-f18}).

\hspace{0.5cm}On the other hand, by increasing $N$ we are allowing more
general functionals, and thus the bound $f_{N}(d)$ can and does get stronger.
This is intuitively clear since for larger $N$ the Taylor-expanded sum rule
includes more and more constraints.

\hspace{0.5cm}Compared to the results of \cite{R}, the bound on the anomalous
dimension $\Delta_{\min}-2$ is improved by $\sim30\div50\%$ in the range
$1\leq d\leq1.7$ that we explored.

\item We have pushed our analysis to such large values of $N$ in the hope of
seeing that the bound saturates as $N\rightarrow\infty$. Indeed, we do observe
signs of convergence in Figs.~\ref{figN},\ref{figNlog}, especially at
$d\gtrsim1.1$. In fact, we have observed that the bounds $f_{N}(d)$ starting
from $N=8$ follow rather closely the asymptotic behavior%
\[
f_{N}(d)\simeq f_{\infty}(d)+\frac{c(d)}{N^{2}},\quad(1\leq d\leq1.7).
\]
An approximation to the optimal bound $f_{\infty}(d)$ can thus be found by
performing for each $d$ a fit to this formula. This approximation is shown by
a dashed line in Fig.~\ref{figN}. From this rough analysis we conclude that
the optimal bound on the anomalous dimension $\Delta_{\min}-2$ is probably
within $\sim10\%$ from our current bound.

\item We have $f_{N}(d)\rightarrow2$ continuously as $d\rightarrow1.$ The
point $d=1,~\Delta_{\min}=2$ corresponds to the free scalar theory.

\hspace{0.5cm}We don't know of any unitary CFTs that saturate our bound at
$d>1$, see the discussion in Section 6 of \cite{R}. We know however a family
of unitary 4D CFTs in which $\Delta_{\min}=2d$ and which are consistent with
our bound (the red dotted line in Fig.~\ref{figN}). This \textquotedblleft
generalized free scalar\textquotedblright\ theory is defined for a fixed $d$
by specifying the 2-point function%
\[
\left\langle \phi(x)\phi(0)\right\rangle =|x|^{-2d}~,
\]
and defining all other correlators of $\phi$ via Wick's theorem. This simple
procedure gives a well-defined CFT, unitary as long as $d\geq1$, which can be
described by a nonlocal action%
\[
S\propto\int d^{4}x\,\phi(\partial^{2})^{d}\phi\,.
\]
The full operator content of this theory can be recovered by studying the OPE
$\phi\times\phi$. In particular, the leading scalar in this OPE has dimension
$2d.$\footnote{This theory can also be realized holographically by considering
a free scalar field of a particular $d$-dependent mass in the AdS geometry and
taking the limit in which 5D gravity is decoupled. We are grateful to Kyriakos
Papadodimas for discussions about the generalized free scalar CFT.}
\end{enumerate}

\subsection{2D analogue}

\label{sec:2Dpreview}

Although our main interest is in the 4D CFTs, our methods allow a parallel
treatment of the 2D case. The main characteristics of the 2D situation were
described in Section 6.1 of \cite{R}, here will briefly review them.

\begin{enumerate}
\item At present we can only take advantage of the finite-dimensional
$SL(2,\mathbb{C})$ symmetry and not of the full Virasoro algebra of the 2D
CFTs. In particular, our results are independent of the 2D central charge $c$.

\item The unitarity bounds for\ $SL(2,\mathbb{C})$ primaries\footnote{Known as
quasi-primaries in 2D CFT literature.} in 2D have the form%
\[
\Delta\geq l,\quad l=0,1,2\ldots,
\]
where $l$ is the Lorentz spin.

\item The $SL(2,\mathbb{C})$ conformal blocks in 2D are known explicitly
\cite{do1}:%
\begin{equation}
g_{\Delta,l}(u,v)=\frac{(-)^{l}}{2^{l}}\left[  \,f_{\Delta+l}(z)f_{\Delta
-l}(\bar{z})+(z\leftrightarrow\bar{z})\right]  . \label{eq-2D}%
\end{equation}

\end{enumerate}

Using the unitarity bounds, the known conformal blocks, and the sum rule
(\ref{eq-F}), valid in any dimension, we can try to answer the same question
as in 4D. Namely, for a $SL(2,\mathbb{C})$ scalar primary $\phi$ of dimension
$d$, what is an upper bound on the dimension $\Delta_{\min}$ of the first
scalar operator appearing in the OPE $\phi\times\phi$? I.e. we want a 2D
analogue of the bound~(\ref{eq-bound}). Since the free scalar is dimensionless
in 2D, the region of interest is $d>0$.

Fig.~\ref{figN2D} summarizes our current knowledge of this bound:\footnote{See
Appendix B for the results in tabular form.}

\begin{itemize}
\item The dotted line is the old $N=2$ bound presented in \cite{R}. The solid
line is the $N=12$ improved bound obtained by us.\footnote{We are grateful to
Erik~Tonni for providing us with the large $\Delta,l$ asymptotics of the 2D
conformal block expansion coefficients, necessary to obtain this bound.}. An
numerical fit to this bound is given by:%
\[
f_{12}^{(2D)}(d)\simeq\left\{
\begin{array}
[c]{ll}%
4.3d+8d^{2}-87d^{3}+2300d^{4},\quad & d\lesssim0.122\,,\\
0.64+2.87d, & d\gtrsim0.122\,.
\end{array}
\right.
\]
Clearly, the improvement compared to \cite{R} is significant.

\hspace{0.5cm}It is interesting to note that in 2D we have observed
a much faster convergence for increasing $N$ than in 4D. In fact,
already with $N=6$ it is possible to obtain a bound rather close to
the one shown in Fig.~\ref{figN2D}, although with a slightly rounded
\textquotedblleft knee\textquotedblright. We have also computed
several points for $N=16$ and haven't seen much improvement.

\item The dashed line and scattered crosses correspond to various OPEs
realized in explicit examples of exactly solvable unitary 2D CFTs (minimal
models and the free scalar theory), see \cite{R}. They all respect our bound.
\end{itemize}

%

\begin{figure}
[ptbh]
\begin{center}
\includegraphics[
height=3.0493in,
width=4.5792in
]%
{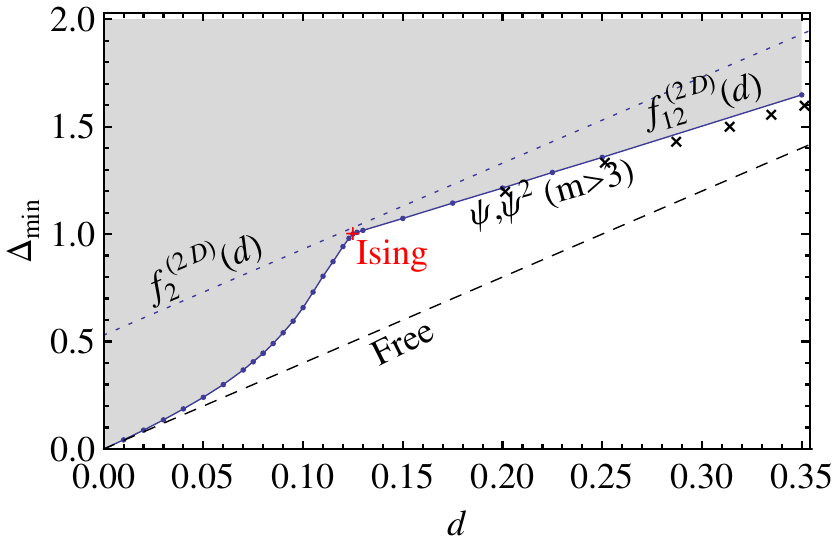}%
\caption{\textit{See the text for an explanation. The red cross denotes the
position of the Ising model, the black crosses marked }$\psi,\psi^{2}$\textit{
correspond to the OPEs realized in the higher minimal models, as in Fig.~15 of
\cite{R}. The shaded region is excluded.}}%
\label{figN2D}%
\end{center}
\end{figure}

It is instructive to compare this plot with its 4D counterpart,
Fig.~\ref{figN}. While we do not know of any CFTs saturating the 4D bound, the
2D unitary minimal models $\mathcal{M}(m,m+1)$, $m=3,4,\ldots$, contain the
OPEs%
\begin{equation}
\psi\times\psi=1+\psi^{2}+\ldots,\quad\Delta_{\psi}=\frac{1}{2}-\frac
{3}{2(m+1)},\quad\Delta_{\psi^{2}}=2-\frac{4}{m+1}, \label{eq:psi}%
\end{equation}
which come quite close to saturating the 2D bound.

More precisely, our 2D bound starts at $(0,0)$ tangentially to the line
$\Delta=4d$ realized in the free scalar theory, then grows monotonically and
passes remarkably closely above the Ising model point $(\Delta_{\sigma}%
,\Delta_{\varepsilon})=\left(  1/8,1\right)  $. After a \textquotedblleft
knee\textquotedblright\ at the Ising point, the bound continues to grow
linearly, passing in the vicinity of the higher minimal model points
(\ref{eq:psi}).

It is curious to note that if we did not know beforehand about the Ising
model, we could have conjectured its field dimensions and the basic OPE
$\sigma\times\sigma=1+\varepsilon$ based on the singular behavior of the 2D
bound at $d=1/8.$

On the other hand, nothing special happens with the 2D bound at the higher
minimal model points, it just interpolates linearly in between\footnote{The
straight line fitting the bound would cross the dashed Free theory line just
above $d=0.5$, which is the accumulation point of the minimal models
$\mathcal{M}(m,m+1)$. For larger values of $d$ we expect that the bound
modifies its slope and eventually asymptotes to the Free line.}. Most likely,
this does not mean that there exist other unitary CFTs with intermediate
operator dimensions. Rather, this behavior suggests that the single conformal
bootstrap equation used to derive the bound is not powerful enough to fully
constrain a CFT.

In comparison, it is a bit unfortunate that the 4D bound does not
exhibit any singular points which would immediately stand out as CFT
candidates. Nevertheless, if we assume that the shape of the 4D
bound is a result of an interpolation between existing CFTs (as it
is the case in 2D), we may conjecture that the upward convex
behavior of the functions $f_{N}(d)$ in Fig.~\ref{figN} is due to
the presence of a family of points satisfying the sum rule that can
correspond to exact CFTs. This observation, though speculative,
shows how the presented method can provide a guideline in the study
of 4D CFTs.

\section{Future research directions}

\label{sec:outlook}

The results of this paper and of \cite{R} open up many interesting research
directions, which we would like to list here.

First, there are several important problems in 4D Conformal Field Theory which
can be analyzed by our method and its simple modifications. For example:

\begin{enumerate}
\item One should be able to derive a generalization of our bounds in the
situation when the CFT has a global symmetry, and we are interested in the
lowest dimension \textit{singlet} appearing in the OPE. This is going to have
phenomenological implications by constraining the so-called conformal
technicolor scenarios of ElectroWeak Symmetry Breaking \cite{luty}. This
connection was extensively discussed in \cite{R}.

\item One should be able to derive model-independent bounds on the size of OPE
coefficients. This is going to be relevant for discussions of `unparticle
self-interactions' \cite{Feng}, in the context of unparticle physics scenarios
\cite{me-too-physics}.
\end{enumerate}

Second, the method can also be used in 2D Conformal Field Theory, as was
already demonstrated in Section \ref{sec:2Dpreview}. The main interest here
lies in potential applications to string theory. We will now briefly describe
two such applications.

Physical states of (super)string theory are in 1-1 correspondence with
Virasoro primary operators of a 2D CFT living on the string worldsheet. The
mass of a string state (in string units) is related to the corresponding
primary operator dimension $\Delta$ via%
\[
m^{2}=\Delta-2\,.
\]
We are considering closed string theory for concreteness. When strings
propagate in flat space, the CFT is solvable and the full spectrum of operator
dimensions is known. Realistic string constructions require compactifications
of extra dimensions. In some examples, such as toroidal compactifications, the
CFT is still solvable. In others, such as superstring compactifications on a
generic Calabi-Yau three-fold, the CFT cannot be solved exactly. All what is
generally known is the spectrum of the massless states, which can be obtained
in the supergravity approximation. Of course we expect the massive string
states to be always present, but just how heavy can they be? We know from the
experience with toroidal compactifications that it is impossible to completely
decouple the massive states: as the compactification radius $R\rightarrow0,$
the Kaluza-Klein states become more massive, but the winding modes come down.
Clearly, massive string states are crucial for the consistency of the theory.
What exactly are they doing? A partial answer may be that without their
presence, 4-point functions of the massless state vertex operators would not
be crossing-symmetric. If this intuition is right, it could be used to obtain
model-independent bounds on the lightest massive states in string
compactifications, generalizing the well-known bounds valid for toroidal
compactifications. A similar in spirit general prediction of string gravity,
although in a different context and by using different methods, was obtained
recently in \cite{hellerman}.

When working towards results of this kind, it may be necessary to
generalize our methods so that the information about the 2D CFT
central charge, which is fixed in string theory, can be taken into
account. In practice, one needs an efficient method to evaluate the
full Virasoro conformal blocks. While no closed-form expression as
simple as Eq.~(\ref{eq-2D}) is known, Zamolodchikov's expansion (see
\cite{ZZ})\ can probably be applied.

Finally, as mentioned above in the 4D context, it should be possible to derive
model-independent bounds on the OPE coefficients. Such results must be
accessible via a simple modification of our method, in particular the full
Virasoro conformal blocks are not needed here. One can then apply such bounds
to the dimension 2 operators corresponding to the massless string states (in
an arbitrary compactification). Via the usual dictionary, this would then
translate into general bounds on the tree-level coupling constants in the
low-energy string effective actions.

\section{Summary}

\label{sec:summary}

Prime principles of Conformal Field Theory, such as unitarity, OPE, and
conformal block decomposition, imply the existence of an upper bound $f(d)$ on
the dimension $\Delta_{\min}$ of the leading scalar operator in the OPE
$\phi\times\phi$, which depends only on $\phi$'s dimension $d$.

Moreover, there is an efficient method which allows numerical determination of
$f(d)$ with arbitrary desired accuracy. The method is based on the \textit{sum
rule}, a function-space identity satisfied by the conformal block
decomposition of the 4-point function $\left\langle \phi\phi\phi
\phi\right\rangle $, which follows from the crossing symmetry constraints. In
practical application of the method the sum rule is Taylor-expanded: replaced
by finitely many equations for the derivatives up to a certain order $N$. The
bound $f(d)$ improves monotonically as more and more derivatives are included.
\ In \cite{R}, where the above paradigm was first developed, we numerically
computed the bound for $N=6$.

The present paper extended the study of \cite{R} to higher $N.$ The
goals were to improve the bound, and perhaps to approach the
best-possible bound in case a convergence of the bound is observed.

Our analysis went up to $N=18$, see Fig.~\ref{figN}, and we have
achieved both goals. First, in the range $1\leq d\leq1.7$ that we
explored, the bound on the anomalous dimension $\Delta_{\min}-2$ is
improved by $30\div50\%$ compared to the results of \cite{R}.
Second, we do observe signs of convergence of the bound. We believe
that our current results are close (within $\sim10\%$) to the best
ones achievable with this method.

The results of this paper and of \cite{R} suggest several interesting research
directions, connected with phenomenology and, via the 2D analogue of our
method, with string theory (see Section \ref{sec:outlook}).

\section{Acknowledgements}

We are grateful to J.~Maldacena, A.M.~Polyakov and N.~Seiberg for discussions
of possible applications of our methods in string theory, to A.M.~Polyakov for
bringing the subtleties of the analytic structure of the conformal blocks to
our attention, and to K.~Papadodimas for discussions of the generalized free
scalar theory. We are especially grateful to our collaborators R.~Rattazzi and
E.~Tonni for many discussions related to this project, and in particular to
E.~Tonni for providing us with the large $\Delta,l$ asymptotics of the 2D
conformal block expansion coefficients. This work is partially supported by
the EU under RTN contract MRTN-CT-2004-503369, by MIUR under the contract
PRIN-2006022501, and by the Swiss National Science Foundation under contract
No. 200021-116372. V.R.\thinspace thanks Laboratoire de Physique Th\'{e}orique
de l'Ecole Normale Sup\'{e}rieure for hospitality.

\appendix

\section{Details about numerical algorithms}

\label{sec:numerics}

We now discuss in more detail the issues introduced in Section
\ref{sec:positive}, namely how we can find in practice a linear
functional $\Lambda\lbrack F]$ of the form (\ref{eq-Lambda})
satisfying the positivity property (\ref{eq-ineq}). We will first
describe the general procedure and how it can implemented in a
computer code, and then mention possible algorithmic improvements
and shortcuts that we found useful in our analysis.

Given the complexity of the functions $F_{d,\Delta,l}$, the search
for a positive functional is too hard a task to be attacked
analytically. As already mentioned, we reduce the complexity of the
problem by looking for a functional which is a linear combination of
derivatives up to a given order $N$. The derivative are taken w.r.t.
the selfdual point $X=Y=0$, since the sum rule is expected to
converge fastest around this point and, in addition, the functions
$F_{d,\Delta,l}(X,Y)$ are even in both arguments. The choice of the
functional (\ref{eq-Lambda}) simplifies our task enormously since we
can now work in a finite dimensional space, and the only information
concerning $F_{d,\Delta,l}$ that we need are their derivatives up to
a certain order. Put another way, the F-functions are now considered
as elements not of a function space but of a finite-dimensional
vector space $\mathbb{R}^{s}$, $s=N(N+6)/8$.

The sum rule (\ref{eq-F}) in this picture represents a constraints
on these vectors that, in any CFT, must sum to zero. This
interpretation is discussed in details in \cite{R}. Here we adopt an
equivalent point of view in terms of the dual space of linear
functionals defined on $\mathbb{R}^{s}$ since we find this
prospective closer to the method used to obtain numerically
$\Lambda\lbrack F]$.

Let us fix the notation. We define the $s$-dimensional vector of
Taylor coefficients:
\begin{align}
\ \mathcal{F}_{0}[d,\delta,l]  &  \equiv\left\{  \frac{1}{m!n!}F_{d,\Delta
,l}^{(m,n)}\,|\,m,n\text{ even,~}2\leq m+n\leq N\right\}
\,\,,\label{eq:vector}\\
F_{d,\Delta,l}^{(m,n)}  &  \equiv\partial_{X}^{m}\partial_{Y}^{n}%
F_{d,\Delta,l}|_{X=Y=0},\qquad\delta\equiv\Delta-l-2,\nonumber
\end{align}
and the same vector normalized to the unit length:%
\begin{equation}
\mathcal{F}[d,\delta,l]\equiv\frac{\mathcal{F}_{0}}{\left\Vert \mathcal{F}%
_{0}\right\Vert }\,\,, \label{eq:vector1}%
\end{equation}
where the norm $\left\Vert \mathcal{F}_{0}\right\Vert $ is the usual Euclidean
length of the vector $\mathcal{F}_{0}$.

We form the vector $\mathcal{F}_{0}$ out of the Taylor coefficients
of the function $F_{d,\Delta,,l}$ rather then of its derivatives,
because this way all elements turn out to have approximately the
same order of magnitude, which is preferable in the subsequent
numerical computation. The definition of the normalized vector
$\mathcal{F}$ serves the same purpose. Indeed, as explained in the
following, our numerical analysis consist in finding a solution of a
system of linear inequalities where the coefficient are given by the
elements of $\mathcal{F}[d,\delta,l]$. The solution is more accurate
and easier to extract if all the coefficient are of the same order
of magnitude. Since the existence of the functional $\Lambda$ is not
affected by these rescalings, we opted for the definition
$($\ref{eq:vector}$),$ (\ref{eq:vector1}).

According to the positivity property we look for a functional which is
strictly positive on all but finitely many vectors $\mathcal{F}[d,\delta,l]$.
Let us fix the dimension $d$ of the scalar $\phi_{d}$. Then each pair
$\Delta\,,l$ identifies the semi-space of $(\mathbb{R}^{s})^{\ast}$ of the
functionals positive-definite on the vectors $\mathcal{F}[d,\delta,l]$; let us
call this open sets $U_{d,\Delta,l}$. With this notation the positivity
property (\ref{eq-ineq}) can be restated in the following way: \textit{If for
fixed }$d$\textit{ and }$\Delta_{\text{min}}$
\begin{equation}
\displaystyle\bigcap_{\substack{\Delta\geq\Delta_{\text{min}}%
,\,l=0\phantom{\,\,-}\\\Delta\geq l+2,\,l=2,4,...}}U_{d,\Delta,l}\neq
\emptyset\,\,, \label{eq-intersection}%
\end{equation}
\textit{then the sum rule cannot be satisfied.} The issue is thus to be able
to check whether the intersection (\ref{eq-intersection}) is non-empty, and to
compute the smallest $\Delta_{\text{min}}$ for which this is the case.

Clearly it is not possible nor needed to check all the values of
$\Delta$ as required by the condition (\ref{eq-intersection}). We
can indeed consider only a finite number of them and check if they
admit the existence of a functional or not. This can be achieved
with a double simplification. First, we consider values of $\Delta$
and $l$ only up to a given maximum value (\textquotedblleft
truncation\textquotedblright), and secondly, we discretize the kept
range of $\Delta$ (\textquotedblleft
discretization\textquotedblright). The truncation does not produce a
loss of information since we take into account the large $\Delta$
and $l$ contributions using the asymptotic expressions computed in
Appendix D of \cite{R}. The discretization step requires special
care, see below.

We used \textsc{Mathematica 7} to perform the computations. The algorithm to
extract the smallest value of $\Delta_{\text{min}}$ proceeds in several steps:

\begin{enumerate}
\item Setting up an efficient procedure to compute vectors $\mathcal{F}%
[d,\delta,l].$

\item Selection of the $l$'s and $\delta$'s to be used in checking the
positivity property (\ref{eq-intersection}). For concreteness we report here
the range of $l,\delta$ that we were including:\footnote{In some cases
$2<l<200$ was needed to obtain a functional which would later pass the
positivity check on the non-included values of $l$, see below.}
\begin{align}
\  &  2\leq l\leq l_{\max}=50\,:\qquad0\leq\delta\leq
200\,\,,\label{eq-interval}\\
\  &  l=0:\qquad\qquad\delta_{\min}\equiv\Delta_{\min}-2\leq\delta
\leq200.\nonumber
\end{align}
For each $l$ the range of $\delta$ was discretized, and a discrete
set of points was chosen, called $\Gamma_{l}$ below. The derivatives
of the F-functions approach zero as $\delta\rightarrow\infty$ and
reach the asymptotic behavior for sufficiently large values. We take
a finer discretization where the function are significantly varying
while we can allow to increase the step in the asymptotic region.
More details are given below.

\item Reduction to a Linear Programming problem. With only a finite number of
equations to check, the determination of the intersection of the
$U_{d,\Delta,l}$ becomes a standard problem of \textit{Linear Programming}
which can be solved in finite amount of time. Hence we look for a solution of
the linear system of inequalities
\begin{align}
\ \Lambda\lbrack\mathcal{F}[d,\delta,l]]  &  \equiv\displaystyle\sum
\tilde{\lambda}_{m,n}\mathcal{F}_{m,n}\geq0\,\,,\label{eq-system}\\
\delta &  \in\Gamma_{l},\quad\,l=0\ldots l_{\max}\,\,\,.\nonumber
\end{align}
Clearly, the coefficients $\tilde{\lambda}_{m,n}$ are related to those
appearing in (\ref{eq-Lambda}) by a trivial rescaling depending on $m,n$:%
\[
\tilde{\lambda}_{m,n}=m!n!\lambda_{m,n}%
\]
Further, the asymptotic behavior of the F-functions (see below) tells us that
for large $\delta$ the inequality is dominated by the $(N,0)$ derivative
\begin{equation}
\ \Lambda\lbrack F_{d,\Delta,l}]\longrightarrow\tilde{\lambda}_{N,0}%
F_{d,\Delta,l}^{(N,0)}\quad\,\,(\delta\gg l\gg 1\text{)},
\end{equation}
hence $\tilde{\lambda}_{N,0}$ needs to be positive. By an overall rescaling of
$\Lambda$ we can always achieve
\begin{equation}
\tilde{\lambda}_{N,0}=1\,,
\end{equation}
which we choose as a normalization condition.

\item Extraction of the smallest $\Delta_{\text{min}}$ for which a positive
functional exists. We begin by selecting two points
$\delta_{\min}=\delta_{1}$ and $\delta_{\min}=\delta_{2}>\delta_{1}$
(see (\ref{eq-interval})) such that we know a priori that in the
first case a positive functional does not exists, while in the
second case it does\footnote{We can choose these points blindly as
$\delta_{1}=0$, $\delta_{2}\gg1$, however prior experience can
suggest a choice closer to the final $\delta_{\text{min}}$}.
Starting from these values we apply the bisection method to
determine the critical $\delta_{\text{min}}$ up to the desired
precision: we test if a functional exists for $\delta_{\min
}=(\delta_{2}+\delta_{1})/2$ and we increase or decrease the
extremes of the interval [$\delta_{1},\delta_{2}]$ depending on the
outcome. The procedure we follow is such that in the end the
critical $\delta_{\text{min}}$ is contained in an interval of
relative width $10^{-3}$, i.e. we terminate if $\delta
_{2}-\delta_{1}\leq10^{-3}\delta_{1}$. The plots and the tables
presented in this work correspond to the upper end $\delta_{2}$ of
the final interval, i.e.\ to the the end for which we have found a
functional.
\end{enumerate}

Let us now come back to the point 1. Although computation of the derivatives
can be carried on by brute force Taylor-expanding the F-functions, we can save
time decomposing the computation in various blocks. From equation (\ref{eq-F})
we see the rather simple dependence on the parameter $d$, which translates in
a polynomial dependence once the function is Taylor-expanded in $X$ and $Y$.
We therefore separately computed the dependence on $d$ once and for all as a
matrix $M(d)_{mn|ij}$. To compute Taylor coefficients of F-functions, this
matrix is contracted with two vectors containing one-dimensional Taylor
coefficients of the function $k_{\beta}(x),$ see (\ref{eq-DO}). The latter
derivatives are pre-computed for several values of $\beta$ with a fine step
and stored. For definiteness we report the interval we used:
\begin{equation}
\ 0\leq\beta\leq10^{2}\qquad\text{step:}\,10^{-3}\,.
\end{equation}
For larger $\beta$ we made use of the analytic expression of the asymptotics
instead of computing the derivatives numerically (see below).

Finally let us discuss the choice of the discretization and the truncation in
$\Delta$ and $l$. This step is of fundamental importance in order to reduce
the time needed to perform the computation.

In Appendix D of \cite{R} it is shown that for large values of $\delta$ and
$l$ the functions $F_{d,\Delta,l}$ approach an asymptotic behavior. We have
checked that outside the range of values (\ref{eq-interval}) we can safely use
the approximate expression
\[
\ F_{d,\Delta,l}^{(m,n)}\sim const.(2\sqrt{2})^{m+n+2}\frac{(l+\delta
)^{m+1}l^{n+1}}{(m+1)(n+1)}\,,\,
\]
For large $l,\delta$ the vector $\mathcal{F}[d,\delta,l]$ is
dominated by the components where $m+n$ assumes the highest allowed
value $N$. Hence we can take into account this large $l,\delta$
behavior imposing additional constraints:
\[
\ \Lambda\lbrack\mathcal{F}_{\theta}]\geq0\,\,,\qquad\mathcal{F}_{\theta
}=\left\{
\begin{array}
[c]{ccc}%
\frac{(\cos\theta+\sin\theta)^{m+1}\cos\theta^{n+1}}{(m+1)(n+1)} & \text{if} &
m+n=N\\
0 &  & \text{otherwise}%
\end{array}
\right.  \,\,,\qquad\tan\theta\equiv\frac{\delta}{l}\,,
\]
where we have dropped irrelevant positive constants not depending on $m\,,n$.

Now comes the discretization: in the range of values
(\ref{eq-interval}), as well as in the interval
$\theta\in\lbrack0,\pi/2]$, we can allow to take only a discrete,
finite number of points. For $\theta$ we take a fixed small step.
However, for $\delta$ we try to concentrate the points in the region
where the unit vector $\mathcal{F}[d,\delta,l]$ is significantly
varying. A measure of
this is given by the norm of its derivative w.r.t. $\delta$%
:\footnote{\label{note-fd}In practice the derivative $\partial/\partial\delta$
is evaluated by using the finite-difference approximation.}%
\[
\ \mathcal{N}=\left\Vert \frac{\partial}{\partial\delta}\mathcal{F}%
[d,\delta,l]\right\Vert \,.
\]
We discretize by taking the spacing between two consecutive values
of $\delta$ equal $c/\mathcal{N}$, where $c$ is a small fixed number
($c=0.02\div0.05$ was typically taken in our work). Clearly when the
unit vector is slowly varying the discretization step is large,
while it is refined where it is changing rapidly, and where
presumably more information is encoded. Typically we get about a
hundred $\delta$ values for each $l,$ but only a few dozen of those
above $\delta>50$.

The sets $\Gamma_{l}$, one for each $l$, of values of $\delta$
obtained in this way are the ones referred to at point 2 above. In
constructing the linear system that we use at point 3 we consider
additional intermediate points between two subsequent $\delta$'s. In
order to understand why we do this, let us assume that we have found
a functional $\Lambda$ which is positive for all the values of
$\delta$ contained in $\Gamma_{l}$. Since we considered a discrete
set of values, it may and actually does happen that for intermediate
values of $\delta$ (which were not included in $\Gamma_{l}$) the
functional becomes slightly negative. In \cite{R} this issue was
solved looking for solution of the form $\Lambda\lbrack
F_{d,\Delta,l}]>\varepsilon$, so that for intermediate values this
condition could be violated but the positivity was safe. In the
current work we found it more convenient to build the linear system
in the following way:

\begin{itemize}
\item for each $\delta\in\Gamma_{l}=\{\delta_{1},....,\delta_{i},\delta
_{i+1},....\}$ we evaluate the vector $\mathcal{F}[d,\delta,l]$.

\item for any two consecutive points $\delta_{i},\delta_{i+1}$, we consider
the first-order Taylor expansion of the vector $\mathcal{F}[d,\delta,l]$
around $\delta=\delta_{i}$ and evaluate it at half-spacing between $\delta
_{i}$ and $\delta_{i+1}$:$^{\text{\ref{note-fd}}}$
\begin{equation}
\ \mathcal{F}_{1/2}[d,\delta_{i},l]\equiv\mathcal{F}[d,\delta_{i},l]+\left(
\frac{\delta_{i+1}-\delta_{i}}{2}\right)  \frac{\partial}{\partial\delta
}\mathcal{F}[d,\delta_{i},l] \label{eq:additional point}%
\end{equation}
and we add the constraints $\Lambda\lbrack\mathcal{F}_{1/2}]\geq0$ to the
linear system (\ref{eq-system}).
\end{itemize}

\begin{figure}
[ptbh]
\begin{center}
\includegraphics[
height=1.8421in,
width=5.5919in
]%
{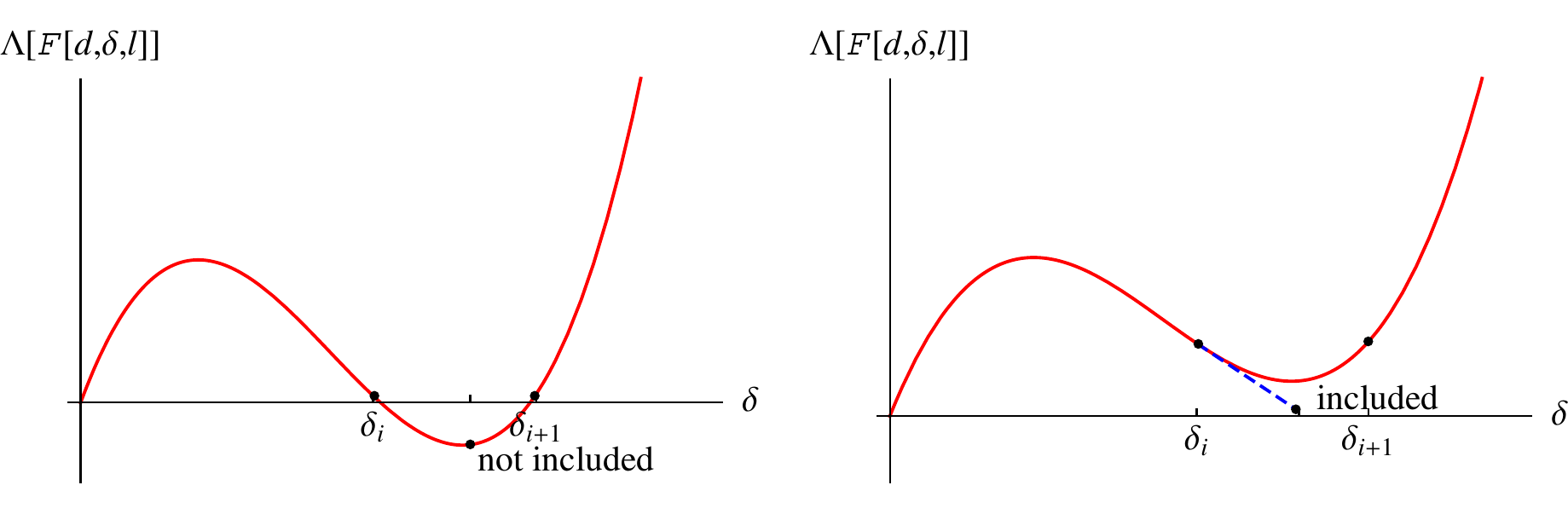}%
\caption{\emph{Imposing the positivity of the functional on a discrete set of
points, it could happens that the intermediate points don't satisfy
$\Lambda\lbrack\mathcal{F}[d,\delta,l]]\geq0$ (on the left). However adding
the constraint $\Lambda\lbrack\mathcal{F}_{1/2}[d,\delta_{i},l]]\geq0$, see
(\ref{eq:additional point}), we can be sure that the functional is positive on
all the neglected points (on the right).}}%
\label{fig:incl}%
\end{center}
\end{figure}

These additional constraints become important to keep the functional positive
near the $\delta$'s for which the inequalities $\Lambda\lbrack\mathcal{F}%
]\geq0$ are close to saturation, while they are redundant away from those
points. Indeed, assume that for some $\delta_{i}$ and $\delta_{i+1}$ the
functional is exactly vanishing. Then at the intermediate point the functional
would be strictly negative, which is not allowed. However, in the presence of
the additional constraint $\Lambda\lbrack\mathcal{F}_{1/2}]\geq0$ this cannot
happen, since $\Lambda\lbrack\mathcal{F]}$ is generically a convex function of
$\delta$ near the minimum. See Figure \ref{fig:incl} for an illustration. Thus
we can be certain that the found functional will be positive also for those
$\delta$ which were not included into $\Gamma_{l}$.%

This certainty has a price. Namely, the opposite side of the coin is that the
added $\mathcal{F}_{1/2}$ constraints are somewhat \textit{stronger} than
needed, and the bigger the discretization parameter $c$, the bigger the
difference. As a result, the found critical value of $\Delta_{\min}$ will be
somewhat \textit{above} the optimal critical value, corresponding to
$c\rightarrow0$. This observation explains why the curves in Figs.~\ref{figN}%
,\ref{figNlog} have small irregularities in the slope. These irregularities
could be decreased by decreasing the value of $c$.

Several comments concerning the numerical accuracy are in order. The
components of the vector $\mathcal{F}[d,\delta,l]$ have been computed using
standard double-precision arithmetic (16 digits). As a consequence all the
numerical results must be rounded to this precision. In particular, quantities
smaller that $10^{-16}$ are considered zero.

In addition, the built-in \textsc{Mathematica 7} function
\texttt{LinearProgramming}, which we used, has an undocumented
\texttt{Tolerance} parameter. Most of the computations were done with
\texttt{Tolerance }equal\texttt{ }$10^{-6}$ (default value). However for
$N=16$ and $N=18$, and for $d<1.1$, we found that \texttt{LinearProgramming
}terminates prematurely, concluding that no positive linear functional exists,
even for some values of $\delta_{\min}$ for which a positive functional for
smaller $N$ was in fact found. The problem disappeared once we set
\texttt{Tolerance} to a lower value ($10^{-12}$). In our opinion,
\texttt{Tolerance} is probably the so-called \textit{pivot tolerance}, the
minimal absolute value of a number in the pivot column of the Simplex Method
to be considered nonzero. Recall that a nonzero (actually negative) pivot
element is necessary in each step of the Simplex Method \cite{lp}. This
interpretation explains why the above problem could occur, and why it could be
overcome by lowering \texttt{Tolerance}.

As described above, our numerical procedure has been designed to be robust
with respect to the effects of truncation and discretization. In addition, for
each $d$, we have tested the last found functional (i.e.~for $\delta_{\min}$
at the upper end $\delta_{2}$ of the final interval $[\delta_{1},\delta_{2}]$)
on the much bigger set of $\delta,l$:
\begin{align}
\  &  2\leq l\leq500\,\,,\quad0\leq\delta\leq500\,\,,\text{step}%
=0.1\,,\nonumber\\
\  &  l=0\,\,,\quad\delta_{\min}\leq\delta\leq500\,\,,\text{step}=0.1\,,
\end{align}
and found that indeed $\Lambda\lbrack\mathcal{F}]\geq0$, within the declared
$10^{-16}$ accuracy.

Finally, we have checked that in all cases the found functionals $\Lambda$ are
such that the inequality $\Lambda\lbrack\mathcal{F}]\geq0$ is in fact strict:
$\Lambda\lbrack\mathcal{F}]>0$, for all but finitely many values of $\delta$
and $l.$ Thus they satisfy the requirements stated in Section
\ref{sec:positive}.

\section{Tables}

\label{sec:tables}

Table 1 contains the sequence of 4D bounds $f_{N}(d)$,
$N=6\ldots18$, for a discrete set of points in the interval
$1<d\leq1.7$. Table 2 contains the 2D bound $f_{12}^{(2D)}(d)$ for a
discrete set of points
in the interval $0<d\leq0.35$. Figs.~\ref{figN}%
, \ref{figNlog}, \ref{figN2D} are based on these tables.

A text file with the unrounded versions of Tables 1,2 and the
functionals used to derive these bounds is included in the source
file of this arXiv submission. The {\sc Mathematica} codes can be
obtained from the authors upon request.

\begin{table}[ptbh]
{
\[%
\begin{array}
[c]{l|lllllll}
&  &  &  & f_{N}(d)-2 &  &  & \\
d-1 & N=6 & N=8 & N=10 & N=12 & N=14 & N=16 & N=18\\\hline
&  &  &  &  &  &  & \\[-10pt]%
0.01 & 0.2045 & 0.2025 & 0.1385 & 0.1356 & 0.1006 & 0.08955 & 0.\\
0.02 & 0.3055 & 0.2956 & 0.2086 & 0.1965 & 0.1636 & 0.1485 & 0.1425\\
0.03 & 0.3895 & 0.3676 & 0.2636 & 0.2496 & 0.2126 & 0.1956 & 0.1916\\
0.04 & 0.4646 & 0.4315 & 0.3165 & 0.3006 & 0.2506 & 0.2366 & 0.2286\\
0.05 & 0.5358 & 0.4878 & 0.3665 & 0.3467 & 0.2896 & 0.2786 & 0.2596\\
0.07 & 0.6688 & 0.598 & 0.4576 & 0.4296 & 0.3927 & 0.3627 & 0.3397\\
0.1 & 0.848 & 0.7348 & 0.5847 & 0.5486 & 0.5026 & 0.4705 & 0.4448\\
0.15 & 1.106 & 0.957 & 0.7789 & 0. & 0.6807 & 0. & 0.\\
0.2 & 1.354 & 1.163 & 0.9635 & 0.9118 & 0.8396 & 0.8098 & 0.7757\\
0.25 & 1.597 & 1.362 & 1.142 & 0. & 1.002 & 0. & 0.\\
0.3 & 1.841 & 1.56 & 1.319 & 1.249 & 1.163 & 1.127 & 1.085\\
0.35 & 2.085 & 1.759 & 1.494 & 0. & 1.317 & 0. & 0.\\
0.4 & 2.329 & 1.957 & 1.669 & 1.578 & 1.473 & 1.436 & 1.389\\
0.45 & 2.561 & 2.159 & 1.843 & 0. & 1.63 & 0. & 0.\\
0.5 & 2.785 & 2.365 & 2.046 & 1.907 & 1.786 & 1.745 & 1.693\\
0.55 & 2.994 & 2.571 & 2.198 & 0. & 1.942 & 0. & 0.\\
0.6 & 3.213 & 2.78 & 2.381 & 2.238 & 2.1 & 2.054 & 1.995\\
0.65 & 3.441 & 2.995 & 2.561 & 0. & 2.258 & 0. & 0.\\
0.7 & 3.638 & 3.21 & 2.744 & 2.574 & 0. & 2.366 & 2.297
\end{array}
\]
}\caption{\textit{4D results. The table contains anomalous dimensions, i.e.
}$d-1$\textit{ and }$f_{N}(d)-2$\textit{ are given. A zero entry means that
the bound for this }$d$\textit{ and }$N$\textit{ has not been computed.}}%
\label{tab1}%
\end{table}

\begin{table}[ptbhptbh]
{
\[%
\begin{array}
[c]{l|l}%
d & f_{12}^{(2D)}(d)\\\hline
& \\[-10pt]%
0.01 & 0.04254\\
0.02 & 0.08752\\
0.03 & 0.1356\\
0.04 & 0.1865\\
0.05 & 0.2406\\
0.06 & 0.2998\\
0.07 & 0.3676\\
0.08 & 0.4455\\
0.075 & 0.4055\\
0.08 & 0.4456\\
0.085 & 0.4907\\
0.09 & 0.5408\\
0.095 & 0.5947\\
0.1 & 0.6577\\
0.105 & 0.73\\
0.11 & 0.8037\\
0.115 & 0.8716\\
0.12 & 0.9418\\
0.123 & 0.9803\\
0.125 & 1.001\\
0.127 & 1.008\\
0.13 & 1.017\\
0.15 & 1.073\\
0.2 & 1.214\\
0.25 & 1.357\\
0.35 & 1.647
\end{array}
\]
}  \caption{\textit{2D results.}}%
\label{tab2}%
\end{table}

\end{document}